# The importance of muon information on primary mass discrimination of ultra-high energy cosmic rays

D. Supanitsky[a], A. Tiba[b], G. Medina-Tanco[b], A. Etchegoyen[a], I. Allekotte[c], M. Gómez Berisso[c], V. de Souza[b], C. Medina[a], J. A. Ortiz[b], R. Shellard[d].
*(a) Laboratorio Tandar, Comisión Nacional de Energía Atómica and CONICET, Av. Del Libertador 8250, (1429) Buenos Aires, Argentina*
*(b) Instituto Astronômico e Geofísico, Univ. de São Paulo, Rua do Matão 1226, 05508-900, Sao Paulo, SP, Brasil*
*(c) Instituto Balseiro, Centro Atómico Bariloche/CNEA and Universidad Nacional de Cuyo, (8400) San Carlos de Bariloche, Río Negro, Argentina.*
*(d) Centro Brasileiro de Pesquisas Físicas, Rua Dr. Xavier Sigaud 150 22290-180 Rio de Janeiro, RJ, Brasil*
Presenter: A. Etchegoyen (etchegoy@tandar.cnea.gov.ar), bra-etchegoyen-A-abs2-he14-poster

Several methods can be used to perform statistical inference of primary composition of cosmic rays measured with water Čerenkov detectors as those in use at the Pierre Auger Southern Observatory. In the present work we assess the impact of additional information about the number of muons in the air shower, on the problem of statistical primary mass discrimination. Several tools are studied, including neural networks, principal component analysis and traditional methods in current use in the field. For our case study we use hypothetical plastic scintillators as muon counters, buried at the side and outside the shade of the water Čerenkov tanks. The study is extended to protons, Si and Fe nuclei impinging on an array with two different spacings, 750 and 1500 m and, therefore, suitable to the 1-10 EeV energy range. A prototype of such a detector is under construction.

## 1. Introduction

The very nature of cosmic rays at the highest energies is surrounded by uncertainties. Despite widely assumed to be mainly extragalactic beyond a few EeV, there is certainly a sizable Galactic contribution perhaps as far as 10 EeV. The variation of the composition as a function of energy inside this energy range is of extreme importance to the understanding of both the Galactic and the extragalactic components.

There is a degeneracy associated with the cosmic rays' spectral shape, which can be reproduced by different combinations of Galactic and extragalactic spectra. Additional information must be obtained either from cosmogenic neutrinos or from detailed composition measurements. Depending on the hardness of the injected spectrum at the extragalactic sources and on the structure and intensity of the intergalactic magnetic fields, the second knee, at approximately $10^{17.7}$ eV, may represent either an abrupt change or the start of a smooth transition to a lighter non-galactic component. The exact form in which this transition takes place might also point to the existence of a major additional source of relativistic particles inside our own Galaxy besides supernova remnant shocks. Even an insight into the physical nature of the ankle, a wide spectral feature around 10 EeV, could also benefit from detailed composition studies.

Several parameters carry, in principle, information on the identity of the primary particle and can be used in order to determine cosmic ray abundances at high energies. For surface detectors, for example, the number of muons at a certain distance $r_0$ of the shower core ($N_\mu(r_0)$), the signal size $S(r_0)$, the arrival time profile of shower particles as characterized by rise time ($T_h$), the fall time ($T_f$) and/or the pulse width ($T_w$), as well as the curvature radius ($R$) can be employed. For fluorescence detectors, the shower maximum atmospheric depth ($X_{max}$), shower maximum fluctuations ($\Delta X_{max}$) and elongation rate ($dX_{max}/dE$) are used as discriminators. $X_{max}$ can also be estimated from pure surface measurements, but at a cost in precision.

The Pierre Auger experiment [1] is able to deliver all the above mentioned surface parameters simultaneously, with the exception of $N_\mu$, and due to its hybrid nature it can also measure $X_{max}$ for ~10% of the sample. We use this experiment as a study case.

In this work we analyze the advantages that the addition of muon information could have on ultra-high energy cosmic ray composition determination, both statistically and on an event-by-event basis. We do this by considering the deployment of underground muon counters at the side of existing water Čerenkov tanks. We use state of the art tools for the simulation of air showers and detector response and for event reconstruction.

Furthermore, since Auger has been originally designed to cover the highest energies, it is fully efficient above ~ 3 EeV [2]. This leaves out the astrophysically very interesting region of the second knee at $\sim 10^{17.7}$ eV and a considerable portion of the ankle, where the most energetic Galactic accelerators turn off and the extragalactic component originated at the highest redshifts becomes dominant. Furthermore, since Auger is located in the Southern hemisphere, it has a full view of the Galactic center, which has been claimed as a possible source at ~ 1-3 EeV [3,4]. Therefore, it seems desirable to extend the capabilities of the experiment to lower energies with as much as possible the same technology, in order to enjoy the benefits of existing infrastructure, cross calibration and the accumulated know how of detectors, data acquisition and analysis tools. Consequently, we also analyze a denser array, an infill, with the same water Čerenkov detectors and muon counters already discussed, but with a spacing half of Auger in its present configuration, i.e. 750 m.

## 2. Numerical approach

The present simulations assume a muon counter array comprising 30m$^2$ surface area detectors with a 100% efficiency. Each muon counter is placed close to an Auger water Čerenkov tank and buried underground to prevent contamination from shower electrons and gammas (via pair production). The assumed depth is such that the threshold muon cut is ~0.9 GeV for vertical muons, similar to AGASA (1 GeV), which for normal rock is equivalent to ~1.5 m. The detectors count muons in 10 ns time bins. We are actually making prototypes of these counters with 4.1cm wide × 2m long plastic scintillator strips with a glued WLS optical fiber, closely following the MINOS [5] design. Each strip fiber is channeled to a pixel of an M64 Hamamatsu phototube. This fine segmentation is to prevent muon pile up within the system processing time.

Muons lose a fraction of their energy, mainly by ionization, when they propagate through the soil. Therefore, just a fraction of them reach the detector. We assume that the energy loss is proportional to the muon track length and constant with energy, then the energy of a muon that traveled a distance $x$ through soil is given by $E(x) = E_0 - \alpha \rho x$, where $E_0$ is the initial energy, $\rho = 2.65 \times 10^6$ g m$^{-3}$ is the density of the soil and $\alpha = 2.1 \times 10^{-7}$ GeV m$^2$/g is the fractional energy loss per optical depth [6].

In order to simulate a muon counter array of the type described above, we modified the program SDSim v3r0 [7], which was originally written to simulate the Pierre Auger array. We use Aires 2.6.0 with QGSJET01 to simulate extensive air showers which are the input of the program.

The shower reconstruction was done by using the standard Auger reconstruction package (CDAS Erv2r4 [8]). The reconstruction of shower direction, core position, and energy were obtained from the Čerenkov detector array rather than from the muon counters. With these three parameters fixed, we proceeded to fit the following lateral distribution function to the detected number of muons at each active counter [9]:

$$N_\mu(r) = P_0 \left(\frac{r}{r_0}\right)^{-\alpha} \left(1+\frac{r}{r_0}\right)^{-\beta} \left(1+\left(\frac{r}{10\,r_0}\right)^2\right)^{-\gamma} \qquad (1)$$

where *r* is core distance. We kept fixed $\alpha = 0.75$, $\gamma = 2.93$ and $r_0 = 320$m while $P_0$ and $\beta$ where free fit parameters. A muon density at any core distance may be obtained from this fit in order to perform composition analyses.

## 3. Numerical results and conclusions

The whole end-to-end chain of simulations, from atmospheric shower to event reconstruction, was performed for events at two characteristic energies, 1 and 10 EeV, and zenith angles 30 and 45°.

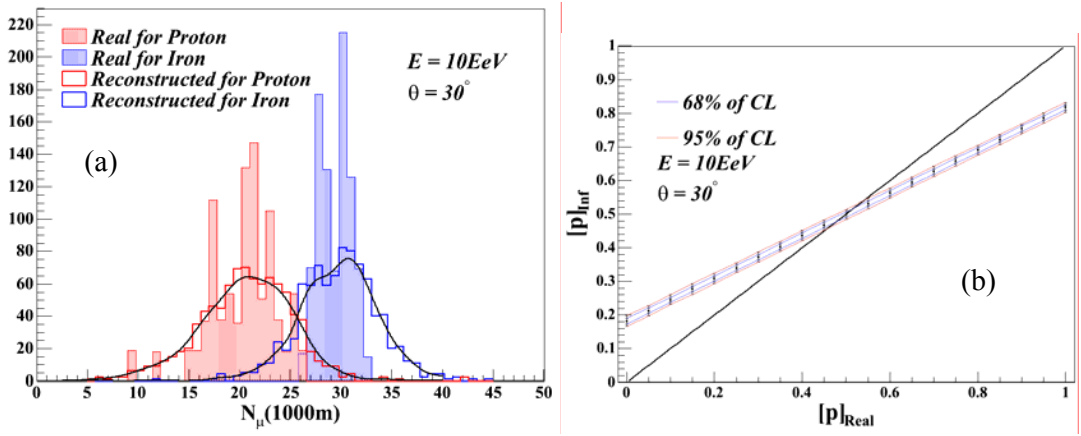

**Figure 1.** (a) Real and reconstructed $N_\mu$ at 1000 m from the shower core. (b) Inferred composition as a function of real composition when the probability distribution function of reconstructed $N_\mu$ is estimated using figure 1.a.

The filled histogram in figure 1.a shows the real distribution of $N_\mu$ at 1000 m from the core (sampled in a 20 m width ring) for a 1500 m spacing. The curves show the same distribution after event reconstruction. These distributions, smoothed in an appropriate way, were used to estimate the probability distribution function of the reconstructed parameter $N_\mu(1000)$ and applied to estimate the composition of independent samples of 1000 events each and actual compositions ranging from 0 to 100%. The bands indicate the 68% and 98% confidence levels (CL).

The same analysis was independently applied to $X_{max}$, $T_h$, $T_f$, $T_w$, $S$ and $R$, and the results are given in figures 2.a-b for the 68 and 98% CL in the abundance determination at different angles and energies. The effectiveness of $N_\mu$ is apparent. Figures 2.c-d show the parameter $\eta = |median(P) - median(Fe)|/\sqrt{\sigma_P^2 + \sigma_{Fe}^2}$, with $\sigma_A^2 = \left(CL_{68}^+ + CL_{68}^-\right)^2$, which is a measure of the power to discriminate primaries on an event-by-event basis. The superiority of $N_\mu$ to all the other parameters taken isolated is again evident. Detector spacing varies with characteristic energy: 750 m at 1 EeV and 1500 m at 10 EeV.

The parameters are correlated to some extent. Nevertheless, better results can still be obtained by applying neural networks. In fact, tests performed training a back-propagation network with Bayesian algorithm [10] show that the absolute error in abundance can be as low as ~2% depending on composition when $N_\mu$ is joined to the remaining surface parameters, down from ~7-10% for hybrids without $N_\mu$. Multivariate techniques such as principal component analysis also go along the same lines supporting these results.

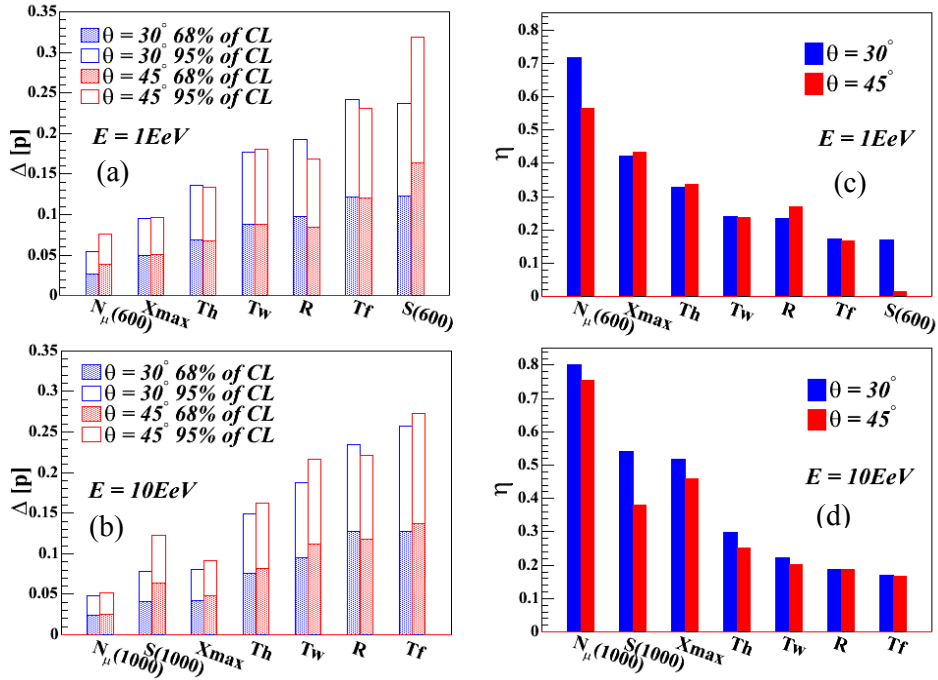

**Figure 2.** Comparison among the various composition determination techniques. The abscise shows the parameters that have been used to study composition at (a-c) 1 and (b-d) 10 EeV. The panels at the left show the confidence levels for the inferred abundance while the right panels are a measure of the capacity to separate individual nuclei.

The present study shows the quantitative advantage for composition analysis in the range 1-10 EeV of the addition of muon information to an array of water Cerenkov detectors. The Pierre Auger Observatory is taken as a study case, both in its present configuration and with the addition of an infill of half the current spacing. Detectors as described by our simulations are being built and will be assembled into a working prototype.

## 4. Acknowledgements

This work is partially supported by CNPq and FAPESP (Brasil) and CONICET, CNEA and IB (Argentina).